\documentclass[epj,dvipsnames]{webofc}
\usepackage[varg]{txfonts}   % Web of Conferences font
\usepackage{hyperref}
\usepackage{color}
\def\VYPx#1#2#3{{\bf #1}, #3 (#2)}  % Volume Year Page, WS-format
\newcommand{\VYP}[4][]{%
  \ifx\\#1\\
    \VYPx{#2}{#3}{#4}% three arguments (1:[{#1}], 4:[{#4}])
  \else
    \href{http://dx.doi.org/#1}{\VYPx{#2}{#3}{#4}}% four arguments  (1:[{#1}], 4:[{#4}])
  \fi
}

\def\etal{ {\it et al.}}
\def\Ttl#1{``{\it #1}'',}

\def\PRD#1#2#3{\href{http://link.aps.org/doi/10.1103/PhysRevD.#1.#3}{Phys.~Rev.~D~\VYP{#1}{#2}{#3}}}

\def\PRL#1#2#3{\href{http://link.aps.org/doi/10.1103/PhysRevLett.#1.#3}{Phys.~Rev.~Lett~\VYP{#1}{#2}{#3}}}

\newcommand{\ApP}[4][]{Astropart.~Phys.\ \VYP[#1]{#2}{#3}{#4}}

\newcommand{\PRep}[4][]{Phys.~Rep.\ \VYP[#1]{#2}{#3}{#4}}

\def\arXiv#1{\href{http://arxiv.org/abs/arXiv:#1}{arXiv:{#1}}}
\def\nucl-th#1{\href{http://arxiv.org/abs/nucl-th/#1}{nucl-th/{#1}}}
\def\beq{\begin{equation}}
\def\eeq{\end{equation} }
\def\bea{\begin{eqnarray}}
\def\eea{\end{eqnarray}}

\def\figref#1{Fig.~\ref{fig:#1}}
\def\figlab#1{\label{fig:#1}}  % Put in caption
\def\eqref#1{Eq.~(\ref{eq:#1})}

\def\eqlab#1{\label{eq:#1}}

\newcommand{\Omit}[1]{}

\newcommand{\BLUE}{\color[named]{Blue}}

\renewcommand{\Ttl}[1]{}

\def\KVI{University of Groningen, KVI Center for Advanced Radiation Technology, 9747 AA Groningen, The Netherlands}
\def\VUB{Vrije Universiteit Brussel, Dienst ELEM, IIHE, Pleinlaan 2, 1050 Brussels, Belgium}

\wocname{ARENA-2016}
\newcommand\wocARENA[1]{\vspace*{-1.1em}\centerline{{\BLUE {\small Contribution [#1] to the proceedings of} ARENA-2016 } %\includegraphics[width=0.4cm,clip]{C:/Users/Olaf/Documents/commissies/ARENA2016/proceedings/LogoARENA}}}
\includegraphics[width=0.4cm,clip]{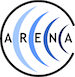}}}
\begin{document}
\title{\wocARENA{58}
Analytic Calculation of Radio Emission from Extensive Air Showers  subjected to Atmospheric Electric Fields}

\author{\firstname{Olaf} \lastname{Scholten}\inst{1,2}\fnsep\thanks{\email{Scholten@KVI.nl}}
\and \firstname{Gia} \lastname{Trinh}\inst{1}
\and \firstname{Krijn D.} \lastname{de Vries}\inst{2}
\and \firstname{Lucas} \lastname{van Sloten}\inst{1}
%\and \firstname{LOFAR} \lastname{CR Key-Science Project}\inst{3}
}

\institute{\KVI
\and       \VUB
%\and       Last address
          }

\abstract{%
We have developed a code that semi-analytically calculates the radio footprint (intensity and polarization) of an extensive air shower subject to atmospheric electric fields.
This can be used to reconstruct the height dependence of atmospheric electric field from the measured radio footprint.
The various parameterizations of the spatial extent of the induced currents are based on the results of Monte-Carlo shower simulations. The calculated radio footprints agree well with microscopic CoREAS simulations.
}
\maketitle
\section{Introduction and summary}
\label{intro}

Atmospheric electric fields during thunderstorm conditions can drastically modify the radio footprint of extensive air showers (EAS). This fact has been used~\cite{Sche15,Tri16} to extract from the measured radio footprint the altitude dependence of the electric fields. In contrast to conventional weather-balloon measurements this offers a non-intrusive way to determine these fields that are essential for understanding lightning initiation and development~\cite{Dwy14}.

The extraction of the atmospheric fields relies on a model calculation of radio emission from an EAS. Accurate codes exists that perform reliable microscopic calculations~\cite{Hue16}. One such code that allows for setting arbitrary atmospheric electric fields is CoREAS~\cite{Hue12} that uses results of a CORSIKA shower evolution. This uses a Monte-Carlo simulation and thus the results of two calculations with very similar but unequal electric fields may differ considerably due to random shower-to-shower fluctuations. As a result it is close to impossible to use CoREAS in a systematic optimization procedure. To resolve this problem we have developed a deterministic semi-analytic calculation that predicts the radio-footprint given a certain altitude dependence of the induced currents in the EAS. An additional advantage is that the analytic code is much less CPU intensive than the microscopic calculation (a few seconds compared to many hours).
As a result it now becomes practical to solve the inverse problem from radio emission, i.e.\ to determine from a measured radio footprint (intensity and polarization distributions on the ground) of a shower the configuration of currents in the atmosphere that produced this.

The code is based on the macroscopic formulation of radio emission as developed in refs~\cite{Scholten08,Wer08,Vries10,Vries11,Sch13} and incorporated in the EVA-code~\cite{Wer12}. An important difference between the EVA-code and the present code (called EVA-light) is that here we use a direct parametrization of the current distribution in EAS. Certain parameters, such as the distribution of the current density as function of distance to the shower axis, are kept fixed, while other parameters related to the altitude dependence of the currents are optimized to fit the data. In addition the new code is optimized for a fast calculation of the complete radio footprint.

Work is still ongoing to refine the parametrizations of the charge-excess and the induced-current clouds where the realistic distributions as determined in Monte-Carlo shower evolution calculations are guiding. Preliminary results show good agreement with those of microscopic CoREAS calculations.

\section{Modelling Radio emission from EAS}

The currents and charges in the EAS are modeled as a charge cloud with a parameterized density profile moving towards Earth with the light velocity. Radio emission is subsequently calculated using the retarded vector potential where we follow the a macroscopic approach to construct the retarded potentials. Furthermore we have put some effort in optimizing the code for the calculation of a complete radio footprint. %Calculate different contributions separately.

In this paper we work with the four-current $j^\mu(t_s,x_s,y_s,h)$ where $z_s=-t_s$ (taking $c=1$) is the distance from the shower front to the ground, called height and always measured along the shower axis, $x_s$, and $y_s$ are the transverse directions, and $h$ is the distance from the shower front to the point in the charge cloud. We use the notation where $\mu=0$ denotes the time (charge) components and $\mu=x,y,z$ the space (current) components of a four vector. The shower front is moving with the light velocity. A particular point in the charge cloud is thus at an height of $\zeta=z_s+h$.

Following the usual notation, the vector potential for an observer at a point $(x_o,y_o,z_o=0)$ on the shower plane, defined as the plane perpendicular to the shower axis going through the point of impact of the shower on the ground, is taken as
\beq
A^\mu(t_o,\vec{x_o})=\int d^3 \vec{x'}\,{j^\mu(t_r,\vec{x'})\over L} \left|{dt_r\over dt}\right|\;,
\eeq
where $L$ is the optical path-length, $L=c(t_o-t_r)$. Only for a homogeneous medium with a constant index of refraction $n$ it is given by $L= nR=n|\vec{x_o}-\vec{x'}|=c(t_o-t_r)$.
Introducing the retarded distance
%- Eq. 2) Misschien is het goed om deze in eerste instantie algemeen te houden tot na Eq. 5). Dus $D= L|dt'/dt|$, waarbij L de optical path length is, dus nog geen nR. Ook in het meest algemene geval van Eq 1) zou dit zo staan. Een andere manier is $g^\mu\nu R_nu V^\mu$ in de teller van Eq. 1 te schrijven, in dit geval hoeft beta=1 zelfs nog niet te gelden.
\beq
{\cal D} \stackrel{\rm def}{=} L \left|{dt \over dt_r}\right|=nR\left.(1-n \vec{\beta}\cdot \hat{n})\right|_{\mbox{ret}}
% \nonumber \\ &=&
 =
 (t-t_r) - n^2\beta(h-\beta t_r)
% \nonumber \\ &=&
 =
 n\sqrt{(-\beta t +h)^2 + (1-\beta^2 n^2)d^2}
 \eqlab{denom} \; ;
\eeq
for a moving point charge with velocity $c\beta$. The vector potential can now be written more compactly as
\beq
A^\mu(t_o,\vec{x_o})=\int d^3 \vec{x'}\,{j^\mu(t_r,\vec{x'})\over {\cal D}} \;.
\eqlab{L-W}
\eeq

The index of refraction depends on the height in the atmosphere through it density dependence as given by the Gladstone-Dale relation, $ n_{\rm GD}=1+n_\rho \;\rho(z) $ where $\rho$ is the air density.
For the evaluation of the retarded distance given in \eqref{denom}, the average value of the index of refraction over the photon trajectory will be used assuming the straight-line approximation for the photon path~\cite{Wer12},
\beq
n = (1/z)\int_0^z(1+n_\rho \rho(\zeta))\,d\zeta \;,
\eeq
which does not depend on the distance of the observer to the shower axis.

To evaluate the integration over $z'$ in \eqref{L-W}, we follow the same approach as used in Ref.~\cite{Wer12} and replace the integral in the $z$-direction by an integral over $\lambda=h_c-h$ where ${\cal D}=0$ at the critical height $h_c$. This substitution allows for an easier calculation of derivatives of the vector potential that are necessary to calculate the electric field since the derivatives vanish on the limits of the $\lambda$ integration. As shown in Ref.~\cite{Wer12}

\subsection{Electric field, charge excess}

The charge excess in the shower front is given by $Q(z)$ propagating with the light velocity in the $-\hat{z}$-direction thus contributing to the zero and the $z$ component of the vector potential.
Substituting the expression for the vector potential, integrating over the spatial extent of the charge cloud and converting to the integration over $\lambda$, the radial component of the electric field can be written as
\beq
E_r^{CX} = -{\partial A^0\over \partial r_o} = -\int dx_s\,dy_s\,{\partial w(r_s)/r_s\over \partial r_s}\, I_\lambda^{CX} \quad ; \quad \quad
I_\lambda^{CX} = \int_0^{h_c} \!\! d\lambda \, f(h_c-\lambda) \left.{Q \over n{\cal D}}\right|_{\lambda=h_c-h} \;,
\eeq
where $r_s^2=(x_s^2+y_s^2)$ is an integration point in the shower front and the position of the observer is denoted with the subscript $o$. The integral is separated into an integration over $\lambda$, a line-sub-shower, along a line parallel to the axis of the full shower, followed by one over the radial extent. In the numerical calculation the results line-sub-showers, $I_\lambda^{CX}(t)$, are stored for a grid of $r_s$ and $d$ values, where the latter is the distance of the observer to line-sub-shower.

In order to obey charge conservation a residue charge line-density $Q'(z) =(dQ/dz)$ is left behind at rest in the atmosphere
%\beq
%A^0_l = \int d^3\vec{x_s}\,\int_{-t_r}^\infty {Q' \over nR} d\zeta \quad ; \quad \quad
%A^3_l = 0 \;, \eqlab{Amu_rst}
%\eeq
which can be ignored as shown in Ref.~\cite{Wer08}.

\subsection{Electric field, Transverse current}

Along a similar line of reasoning as for the charge-excess radiation, the transverse current contribution is written as
\bea
E_x^{TC} &=& -{\partial A^x\over \partial t^o} =\int dx_s\,dy_s {w(r_s)\over r_s} I_\lambda^{TC}  \\
I_\lambda^{TC} &=& \int_0^{h_c}\!\! d\lambda \, f'(h_c-\lambda) \left.{J^x \over n{\cal D}}\right|_{\lambda=h_c-h}  - \int_0^{h_c} \!\! d\lambda \, f(h_c-\lambda) \left.{J^{\prime x} \over n{\cal D}}\right|_{\lambda=h_c-h} \;.
\eea
where $J^{\prime x}=dI^x /dt_r$ and $f'(h)=df/dh$. As before the results for line-sub-showers, $I_\lambda^{TC}(t)$ are calculated once on a grid of $r_s$ and $d$ values to be used subsequently in the calculation of the complete footprint of the electric field.

%\subsection{charge conservation}

Due to the continuity equation we have $d J_x/dx =dQ/dt$, thus for $r_s^2=x_s^2+y_s^2$ we write
$
Q^D_x(t_r,x_s,y_s)= \left[\int_{-\infty}^{t_r} J_x(t) dt\right] \times  {d\, w(r_s)/r_s \over dx_s} $
and a similar term for $J_y$.
This corresponds to a di-pole charge distribution moving with the shower front.  In reality these particles thermalize and will trail well behind the front and thus not contribute to coherent emission. To account for this we include a decrease proportional to $e^{-X/a^D X_0}$,
\beq
\vec{J^D}(t_r)=\int_{-\infty}^{t_r} \vec{J}(t) \,e^{-[X(t_r)-X(t)]/(a^D X_0)}\,dt \;,
\eqlab{JD}
\eeq
where $X$ depends on $t$ through its dependence on the height in the atmosphere. In \eqref{JD} $X_0$ is the electron mean free path and $a^D$ a parameter that is adjusted to get best agreement with microscopic shower calculations. Following a similar notation as before the contribution to the electric field is
\beq
E_x^D = -\int dx_s\,dy_s\,{d\, w(r_s)/r_s \over r_s\, dr_s}\, I_D^x \quad ; \quad \quad
 I^D_x = \int d\lambda \, f(h_c-\lambda) \left. {J^D_x\over n{\cal D}} \right|_{\lambda=h_c-h} \;,
\eeq
where a similar expression is obtained for $E_y^D$.

\subsection{Parameterizations}

The density distribution of the four current is parameterized as
\beq
j^\mu(t,x,y,h)={w(r) \over r} \, f(h,r) \, J^\mu(t)
\quad ; \quad
w(r)=N_w\,r/M_0 (r/M_0 +1)^{-2.5}
\quad ; \quad
f(h,r)=N_f\, {u}^{\alpha} e^{-2u}
\eqlab{DefCloud}
\eeq
where $r=\sqrt{x^2+y^2}$, $u={h/ \alpha\,\lambda}$, $M_0$ the Moliere radius, $h$ the distance behind the shower front. The normalization constants $N_w$ and $N_f$ in functions $w$ and $f$ are determined by $\int w(r)\, dr=1$ and $\int_0^\infty f(h,r)\, dh=1\; \forall\, r$. In this way $J^\mu(t)$ is the charge and current at time $t$ integrated over the whole shower front.
Note that $w(r)$ corresponds to $r$ times the NKG function for $s=2$.
The two parameters in $f$, the pancake thickness $\lambda(r)$ and $\alpha(X)$, depend on distance to the shower axis, $r$, and penetration depth, $X$, as
\beq
\lambda(r)= \max[\lambda_0, \lambda_1 \, (r/r_1) ]
\quad ; \quad
\alpha(X)= 1+0.5\times \sqrt{X/X_{max}} \;.
\eeq
The radial dependence of the pancake thickness is given by $\lambda(r)$ such that for $r=0$ we have $\lambda(0)=\lambda_0$ increasing linearly with distance on an exponential grid where at a distance of $r_1=$100~m from the core we have $\lambda(100)=\lambda_1$.
With increasing $\alpha$ the maximum in the particle density shifts to larger distances behind the shower front.

Optimization of the parameters is still ongoing. We note that choosing $M_0=30$~m, $\lambda_0=0.15$~m, $\lambda_1=3$~m, reproduces the main features of the charge cloud as deduced from microscopic calculation. This also gives rise to a radio footprint that agrees well with that obtained from CoREAS simulations. We are still investigating the extent to which the distributions need to be chosen differently for charge excess and transverse current in view of our findings in Ref.~\cite{Tri16}. We note that changing the pancake thickness with distance to the shower axis is essential to be able to reproduce the (not frequency filtered) pulse shape as obtained from the CoREAS calculations. The penetration-depth dependence of $\alpha$ does not have a strong influence on the calculations.

\Omit{
\subsubsection{Comparison with shower simulations}

The radial dependence of the pancake thickness at the shower max.

\begin{figure}[h]
 \includegraphics[width=0.44\textwidth]{../distributions.pdf}
 \caption{radial dependence in microscopic calculations.\figlab{MCresults}}
\end{figure}

\begin{figure}[h]
 \includegraphics[width=0.44\textwidth]{../sh-pars.pdf}
 \caption{parametrized dependence of the shower parameters.\figlab{AnalyticParam}}
\end{figure}

Comparing the results of CONEX Monte Carlo, shown in \figref{MCresults}, with the parametrization given in \figref{AnalyticParam} one sees the perfect agreement between the two.

Variation of the pancake thickness with shower age.

Radial dependence is crucial to get correct pulse-shape as produced by CoREAS.
}

%\appendix
%\section{Parameterizations of pancake thickness}

%\section{Parameterizations of E-field effect}

%\section{Programming details}

%
% BibTeX or Biber users please use (the style is already called in the class, ensure that the "woc.bst" style is in your local directory)
% \bibliography{name or your bibliography database}
%
% Non-BibTeX users please use
%

\end{document}